\documentclass[journal,draftcls,onecolumn]{IEEEtran}\usepackage{amssymb,amsmath,MnSymbol}
\usepackage{graphicx}
\newcommand{\Z}{\mathbb{Z}}
\newcommand{\R}{\mathbb{R}}

\newcounter{defcompt}
\newenvironment{definition}[1][Definition]{\addtocounter{defcompt}{1} 
\begin{trivlist}
\item[\hskip \labelsep {\bfseries #1 \arabic{defcompt}. }]}{\end{trivlist}}

\title{The relation between Granger causality and directed information theory: a review}

\author{ Pierre-Olivier Amblard$^{1,2}$ and Olivier J.J. Michel$^1$ \\
 $^1$ GIPSAlab/CNRS UMR 5216/ BP46, \\ 38402 Saint Martin d'H\`eres cedex, France \\
 $^2$ The University of Melbourne, Dept. of Math\&Stat. 
 Parkville, VIC, 3010, Australia\\
{\tt bidou.amblard@gipsa-lab.inpg.fr}  \\
{\tt olivier.michel@gipsa-lab.grenoble-inp.fr}
}

\begin{document}
\maketitle
%%%%%%%%%%%%%%%%%%%%%%%%%%%%%%%%%%%%%%%%%%%%%%%%%%%%%%%%%%%%
\begin{abstract}This report reviews the conceptual and theoretical links between Granger causality and directed information theory.  We begin with a short historical tour of Granger causality, concentrating on its closeness to information theory. The definitions of Granger causality based on prediction are recalled, and the importance
of the observation set  is discussed. We present the definitions based on conditional independence. The notion of instantaneous coupling is included in the definitions. The concept of Granger causality graphs is discussed. We  present directed information theory from the perspective of studies of causal influences between stochastic processes. Causal conditioning appears to be the cornerstone for the relation between information theory and Granger causality. In the bivariate case, the fundamental measure is the directed information, which decomposes as the sum of the transfer entropies and a term quantifying instantaneous coupling.
We  show the decomposition of the mutual information into the sums of the transfer entropies and  the instantaneous coupling measure, a relation known for the linear Gaussian case. We study the multivariate case, showing that the useful decomposition is blurred by instantaneous coupling. 
The links are further developed by studying how measures based on directed information theory naturally emerge from 
Granger causality inference frameworks as hypothesis testing. 
\end{abstract}

% Keywords: add 3 to 10 keywords
{\bf keyword:} {Granger causality, transfer entropy, information theory, causal conditioning, conditional independence}

\section{Introduction}

This review deals with the analysis of influences that one system, be it physical, economical, biological or social, for example, can exert over another. 
In several scientific fields, the finding of the influence network  between different systems is crucial. As examples, we can think of 
gene influence networks \cite{RaoHSE06,RaoHSE07}, relations between economical variables \cite{Gran69,Sims72}, communication between neurons or the flow of information between different brain regions \cite{Spor10}, or the human  influence on the Earth climate \cite{KaufS97,Tria01}, and many others.

\begin{figure}[tb]
\begin{center}
\includegraphics[scale=.7]{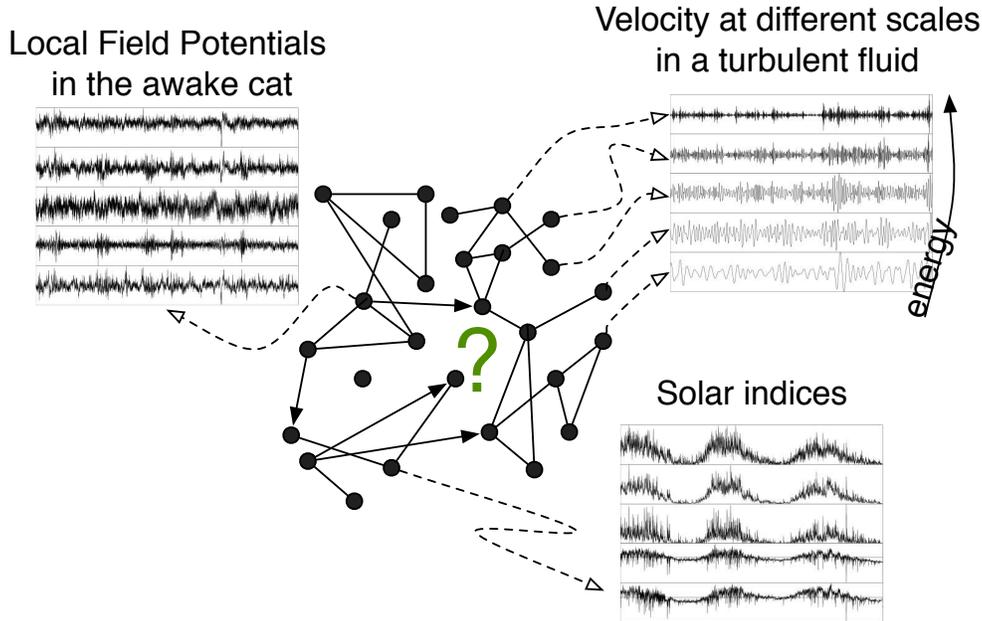}
\end{center}
 \caption{Illustration of the problem of information flow in networks of stochastic processes. Each node of the network is associated to a signal. Edges between nodes stand for dependence (shared information) between the signals. The dependence can be directed or not. This framework can be applied to different situations as solar physics, neuroscience or the study of turbulence in fluids, as illustrated by the three examples depicted here.}
 \label{network:fig}
 \end{figure}

The context studied in this report is illustrated in figure \ref{network:fig}. For a given system, we have at disposal a number of different measurements. In neuroscience, these can be local field potentials recorded in the brain of an animal; In solar physics, these can be solar indices measured by sensors onboard some satellite; In the study of turbulent fluids, these can be the velocity measured at different scales in the fluid (or can be as in the figure, the wavelet analysis of the velocity at different scales).
For these different examples, the aim is to find dependencies between the different measurements, and if possible, to give a direction to the dependence. In neuroscience, this will allow to understand how information flows between different areas of the brain; In solar physics, this will allow to understand the links between indices and their influence on the total solar irradiance received on Earth; In the study of turbulence, this can confirm the directional cascade of energy from large down to small scales.

In a graphical modeling approach, each signal is associated to a particular node of a graph, and dependence are represented by edges, directed if a directional dependence exists. The questions addressed in this paper concern 
the assessment of directional dependence between signals, and thus concern the inference problem of estimating the edge set in the graph of signals considered.

Climatology and neuroscience were  already  given as examples by Norbert Wiener in 1956 \cite{Wien56}, a paper which inspired econometrist Clive Granger to develop what is now termed Granger causality \cite{Gran03}. Wiener proposed in this paper that a signal $x$ causes another time series $y$, if the past of $x$ has a strictly  positive influence on the quality of prediction of $y$. Let us quote Wiener \cite{Wien56}:
\begin{quote}
``As an application of this, let us consider the case where $f_1(\alpha)$ represents the temperature at 9 A.M. in Boston and 
$f_2(\alpha)$  represents the temperature at the same time in Albany. We generally suppose that weather moves
from west to east with the rotation of the earth; the two quantities $1-C$  and its correlate in the other direction will enable us to make a precise statement containing some if this content and then verify whether this statement is true or not. Or again, in the study of brain waves we may be able to obtain electroencephalograms more or less corresponding to electrical activity in different part of the brain. Here the study of coefficients of causality running both ways and of their analogues for sets of more than two functions $f$ may be useful in determining what part of the brain is driving what other part of the brain {\em in its normal activity}."
\end{quote}

In a wide sense, Granger causality can be summed up as a theoretical framework based on conditional independence to assess directional dependencies between time series. It is interesting to note that Norbert Wiener influenced Granger causality, as well as another field dedicated to the analysis of dependencies: information theory. Information theory has led to the definition of quantities that measure the uncertainty on variables using probabilistic concepts.
Furthermore, this has led to the definition of measures of dependence based on the decrease in uncertainty relating to one variable after observing another one. Usual information theory is, however, symmetrical. For example, the well-known mutual information rate between two stationary time series is symmetrical under an exchange of the two signals: the mutual information assesses the undirectional dependence. Directional dependence analysis viewed as an information-theoretic problem requires the breaking of the usual  symmetry of information theory. This was realized in the 1960's and early 1970's by Hans Marko, a German professor of communication. He developed the bidirectional 
information theory in the Markov case \cite{Mark73}. This theory was later generalized by James Massey and Gerhard Kramer, to what we may now call directed information theory \cite{Mass90,Kram98}. 

{\em It is the aim of this report to review the conceptual and theoretical links between Granger causality and directed information theory.} 

Many information-theoretic  tools have been designed for the practical implementation of Granger causality ideas. We will not show all of the different measures proposed, because they are almost always particular cases of the measures issued from directed information theory. Furthermore, some measures
might have been proposed in different fields (and/or at different periods of time) and have received different names. We  will only consider the well-accepted names. This is the case, for example, of `transfer entropy', as coined by Schreiber in 2000 \cite{Schr00}, but which appeared earlier under different names, in different fields, and might be considered under slightly different hypotheses. Prior to developing a unified view of the links between Granger causality and information theory, we will provide a survey of the literature, concentrating on studies where information theory and Granger causality are jointly presented. 

Furthermore, we will not review any practical aspects, nor any detailed applications. In this spirit, this report is different from \cite{HlavSPVB07}, which concentrated on the estimation of information quantities, and where the review is restricted to transfer entropy. For reviews on the analysis of dependencies between systems and for applications of Granger causality in neuroscience, we refer to \cite{Palu07,GourBF06}. We will mention however some important practical points in our conclusions, where we will also discuss some current and future directions of research in the field.

\subsection{What is, and what is not, Granger causality}

We will not  debate the meaning of causality or causation. We instead refer to \cite{Pear00}. However, we must emphasize that Granger causality actually measures a statistical dependence between the past of a process and the present of another. In this respect, the word causality in Granger causality takes on the usual meaning that a cause 
occurs {\it prior } its effect. However, nothing in the definitions that we will recall precludes that signal $x$ can simultaneously be Granger caused by $y$ and be a cause of $y$! This lies in the very close connection between Granger causality and the feedback between times series. 

Granger causality is based on the usual concept of conditioning in probability theory, whereas approaches developed for example in \cite{Pear00,Laur01} relied on causal calculus and the concept of intervention. In this spirit, intervention is closer to experimental sciences, where we imagine that we can really, for example, freeze some system and measure the influence of this action on another process. It is now well-known that causality in the sense of between random variables can  be inferred unambiguously only in restricted cases, such as directed acyclic graph models \cite{Whit89,Laur96,Pear00,Laur01}.
In the Granger causality context, there is no such ambiguity and restriction.

\subsection{A historical viewpoint}

In his Nobel prize lecture in 2003, Clive W. Granger mentioned that in  1959,  Denis Gabor pointed out the work of Wiener to him, as a hint to solve some of the difficulties he met in his work. Norbert Wiener's paper is about the theory of prediction \cite{Wien56}. At the end of his paper, Wiener proposed that prediction theory  could be used to define causality between time series. Granger further developed this idea, and came up with a definition of causality and testing procedures \cite{Gran63,Gran69}.

In these studies, the essential stones were laid. Granger's causality  states that a cause must occur before the effect, and that causality is relative to the knowledge that is available.  This last statement deserves some comment. When testing for causality of one variable on another, it is assumed that the cause has information about the effect that is unique to it; {\it i.e. } this information is unknown to any other variable. Obviously, this cannot be verified for variables that are not known. Therefore, the conclusion drawn in a causal testing procedure is relative to the set of measurements that are available.  A conclusion  reached based on a set of measurements can be altered if new measurements are taken into account. 

Mention of information theory is also present in the studies of Granger. In the restricted case of two  Gaussian signals, Granger already noted the link between what he called the `causality indices' and the mutual information (Eq. 5.4 in \cite{Gran63}). Furthermore, he already foresaw the generalization to the multivariate case, as he wrote in the same paper: 
\begin{quote}
``In the case of $q$ variables, similar equations exist if coherence is replaced by partial coherence, and a new concept of 'partial information' is introduced.''
\end{quote} 
Granger's paper in 1969 does not contain much new information, but rather, it gives a refined presentation of the concepts.

During the 1970's, some studies,  {\it e.g.} \cite{Sims72,CainC75,Hoso77}, appeared that generalized along some of the directions Granger's work, and related some of the applications to economics.
In the early 1980's, several studies were published that established the now accepted definitions of Granger causality \cite{Gran80,FlorM82,Cham82}. These are natural extensions of the ideas built upon prediction, and they rely on conditional independence. Finally, the recent studies of Dalhaus and Eichler allowed the definitions of Granger causality graphs \cite{DahlE03,Eich06,Eich11}. These studies provide a counterpart of graphical models of multivariate random variables to multivariable stochastic processes.

In two studies published in 1982 and 1984 \cite{Gewe82,Gewe84}, Geweke, another econometrician, set up a full treatment of Granger causality testing for the Gaussian case, which included the idea of feedback and instantaneous coupling. In \cite{Gewe82}, the study was restricted to the link between two time series (possibly multidimensional). In this study, Geweke defined an index of causality from $x$ to $y$;  It is the logarithm of the {\it ratio} of the asymptotic mean square error when predicting $y$ from its past only, to the asymptotic mean square error when predicting $y$ from its past and from the past of $x$. 
Geweke also defined the same kind of index for instantaneous coupling, and showed, remarkably, that the mutual information rate between $x$ and $y$ decomposes as the sum of the indices of causality from $x$ to $y$ and from $y$ to $x$ with the index of instantaneous coupling. This decomposition was shown in the Gaussian case, and it remains valid in any case when the indices of causality are replaced by transfer entropy rates, and the instantaneous coupling index is replaced by an instantaneous information exchange rate.
This link between Granger causality and directed information theory was further supported by \cite{AmblM09,BarnBS09} (without mention of instantaneous coupling in \cite{BarnBS09}), and the generalization to the nonGaussian case by \cite{AmblM11} (see also \cite{QuinCKH11} for related results).
However, {\it prior } to these recent studies, the generalization of Geweke's idea to some general setting was reported in 1987, in econometry by Gouri\'eroux {\it et al.} \cite{GourMR87}, and in engineering by Rissannen\&Wax \cite{RissW87}. Gouri\'eroux and his co-workers considered a joint Markovian representation of the
signals, and worked in a decision-theoretic framework. They defined a sequence of nested hypotheses, whether causality was true or not, instantaneous coupling was present or not. They then worked out the decision statistics using the Kullback approach to decision theory
\cite{Kull68}, in which discrepancies between hypotheses are measured according to the Kullback divergence between the probability measures under the hypotheses involved. In this setting, the decomposition obtained by Geweke in the Gaussian case was evidently generalised . In 
\cite{RissW87}, the approach taken was closer to Geweke's study, and it relied on system identification, in which the complexity of the model was taken into account. The probability measures were parameterized, and an information measure that jointly assessed the estimation procedure and the complexity of the model was used when predicting a signal. This allowed Geweke's result to be extended to nonlinear modeling (and hence the nonGaussian case), and provided an information-theoretic interpretation of the tests. Once again, the same kind of decomposition of dependence was obtained by these authors. We will see in section \ref{infotheory:sec} that the decomposition holds due to Kramers causal conditioning. These studies were limited to the bivariate case \cite{GourMR87,RissW87}. 

In the late 1990's, some studies began to develop in the physics community on influences between dynamical systems. A first 
route was taken that followed the ideas of dynamic system studies for the prediction of chaotic systems. To determine if one signal influenced another, the idea was to consider each of the signals as measured states of two different dynamic systems, and then to study the master-slave relationships between these two systems (for examples, see \cite{VanqMAV99,QuiaAG00}). The dynamics of the systems was built using phase space reconstruction \cite{KantS04}. The influence of one system on another was then defined by making a prediction of the dynamics in the reconstructed phase space of one of the processes.  
To our knowledge, the setting was restricted to the bivariate case. A second route, which was also restricted to the bivariate case, was taken and relied on information-theoretic tools. The main contributions were from Palu\v{s} and Schreiber \cite{Schr00,PaluKHS01}, with further developments appearing some years later \cite{KaisS02,PaluV07,FrenP07}. In these studies, the influence of one process on the other was measured by the discrepancy between the probability measures under the hypotheses of influence or no influence. Naturally, the measures defined very much resembled the measures proposed by Gouri\'eroux {\it et. al} \cite{GourMR87}, and used the concept of conditional mutual information. The measure to assess whether one signal influences the other was termed
 {\em transfer entropy} by Schreiber. Its definition was proposed under a Markovian assumption, as was exactly done in \cite{GourMR87}. The presentation by Palu\v{s} \cite{PaluKHS01} was more direct and was not based on a decision-theoretic idea. The measure defined is, however, equivalent to the transfer entropy. 
Interestingly, Palu\v{s} noted in this 2001 paper the closeness of the approach to Granger causality, as per the quotation: 
\begin{quote}
``the [latter] measure can also be understood as an information theoretic formulation of the Granger causality concept.''
\end{quote}
 Note that most of these studies considered bivariate analysis, with the notable exception of \cite{FrenP07}, in which the presence of side information (other 
measured time series) was explicitely considered. 

In parallel with these studies, many others were dedicated to the implementation of Granger causality testing in fields as diverse as climatology  (with applications to the controversial questions of global warming) and neuroscience; see  
\cite{SaitH81,KaufS97,KamiDTB01,Tria01,Eich05,Eich06,GourBF06,MoseSCM06}, to cite but a few.

In a very different field, information theory, the problem of feedback has lead to many questions since the 1950's. We will not review or cite anything on the problem created by feedback in information theory as this is not within the scope of the present study, but some information can be found in \cite{CoveT06}. Instead, we will concentrate on studies that are directly related  to the subject of this review. A major breakthrough was achieved by James Massey in 1990 in a short conference paper \cite{Mass90}.
Following the (lost?) ideas of Marko on bidirectional information theory that were developed in the Markovian case \cite{Mark73}, Massey re-examined the usual definition of what is called a discrete memoryless channel in information theory, and he showed that the usual definition based on some probabilistic assumptions prohibited the use of feedback. He then clarified the definition of memory and feedback in a communication channel. As a consequence, he showed that in a general channel used with feedback, the usual definition of capacity that relies on mutual information was not adequate. Instead, the right measure was shown to be {\em directed information}, an asymmetrical measure of the flow of information. These ideas were further examined by Kramer, who introduced the concept of causal conditioning, and who developed the first applications of directed information theory to communication in networks \cite{Kram98}. After some years, the importance of causal conditioning for the analysis of communication in systems with feedback was realized. Many studies were then dedicated to the analysis of the capacity of channels with feedback and the dual problem of rate-distortion theory \cite{Tati00,TatiM09,VenkP07,Kim08}. Due to the rapid development in the study of networks ({\it e.g.,} social networks, neural networks) and of the afferent connectivity problem, more recently many authors made connections between information theory and Granger causality
\cite{AmblM08,Solo08,AmblM09,AmblM09:arxiv,BarnBS09,AmblM11,QuinCKH11}. Some of these studies were restricted to the Gaussian case, and to the bivariate case. Most of these studies did not tackle the problem of instantaneous coupling. Furthermore, several authors realized the importance of directed information theory to assess the circulation of information in networks \cite{AlkhA08,RaoHSE06,RaoHSE07}.

\subsection{Outline}

Tools from directed information theory appear as natural measures to assess Granger causality. 
Although Granger causality can be considered as a powerful theoretical framework to study influences between signals mathematically,  directed information theory provides the measures to  test theoretical assertions practically. As already mentioned, these measures
are transfer entropy (and its conditional versions), which  assesses the dynamical part of Granger causality, and instantaneous information exchange (and its conditional versions), which assesses instantaneous coupling. 

This review is structured here as follows. We will first give an overview of the definitions of Granger causality. These are presented in a multivariate setting. We go gradually from weak definitions based on prediction, to strong definitions based on conditional independence. 
The problem of  instantaneous coupling is then discussed, and we show that there are two possible definitions for it. 
Causality graphs (after Eichler \cite{Eich11}) provide particular reasons to prefer one of these definitions. Section \ref{infotheory:sec}  introduces  an analysis Granger causality from an information-theoretic perspective. We insist on the concept of causal conditioning, which is at the root of the relationship studied. Section \ref{links:sec} then highlights the links. Here, we first restate the definitions of Granger causality using concepts from directed information theory. Then from of a different point of view, we  show how conceptual inference approaches lead to the  measures defined in directed information theory. The review then closes with a discussion of some of the aspects that we do not present here intentionally, and on some lines for further research. 

\subsection{Notations}

All of the random variables, vectors and signals considered here are defined in a common probability space $(\Omega, {\cal B}, P)$. They take values either in $\R$ or $\R^d$, $d$ being some strictly positive integer, or they can even 
take discrete values. As we concentrate on conceptual aspects rather than technical aspects, we assume that the variables considered are 'well behaved'. In particular, we assume finiteness of moments of sufficient order. We assume that continuously valued variables have a measure that is absolutely continuous with respect to the Lebesgue measure of the space considered. Hence, the existence of probability density functions is assumed. Limits are supposed to exist when needed. All of the processes considered in this report are assumed to be stationary.

We work with discrete time. A signal will generically be denoted as $x(k)$. This notation stands also for the value of the signal at time $k$. The collection of successive samples of the signal,  $x_{k}, x_{k+1},\ldots,x_{k+n}$ will be denoted as $x_{k}^{k+n}$. Often, an initial time will be assumed. This can be 0, 1, or $-\infty$. In any case, if we collect all of the sample of the signals from the initial time up to time $n$, we will suppress the lower index and write this collection as $x^n$. 

When dealing with multivariate signals, we use a graph-theoretic notation. This will simplify some connections with graphical modeling. Let $V$ be an index set of finite cardinality $|V|$.
$x_V=\{x_V(k),  k\in \Z\}$ is a $d$-dimensional discrete time stationary multivariate process for the probability space considered. For $a\in V$,   $x_a$ is the corresponding component of $x_V$. Likewise,  for any subset $A\subset V$, $x_A$ is the corresponding multivariate process $(x_{a_1},\ldots,x_{|A|})$. We say that subsets $A,B,C$ form a partition of $V$ if they are disjoint and if $A \cup B\cup C = V$.
The information obtained by observing $x_A$ up to time $k$ is resumed by the filtration generated by $\{ x_A(l), \forall l\leq k\}$. This is denoted as $x_A^k$.  Furthermore, we will often identify $x_A$ with $A$ in the discussion.

The probability density functions (p.d.f.) or probability mass functions (p.m.f) will be denoted by the same notation as $p(x_A^n)$. The conditional p.d.f. and p.m.f.  are written as $p(x_A^n|x_B^m)$. The expected value is denoted as $E[.],E_x[.]$ or $E_p[.]$ if we want to specify which variable is averaged, or under which probability measure the expected value is evaluated.

Independence between random variables and vectors $x$ and $y$ will be denoted as $x \upmodels y$, while conditional independence given $z$ will be written as $x \upmodels y \mid z$.

%%%%%%%%%%%%%%%%%%%%%%%%%%%%%%%%%%%%%%%%%%%%%%%%%%%%%%%%%%%%

\section{Granger's causality}
\label{Grangercausality:sec}

The early definitions followed the ideas of Wiener: A signal $x$ causes a signal $y$ if the past of $x$ helps in the prediction of $y$.
Implementing this idea requires the performing of the prediction and the quantification of its quality. This leads to a weak, but operational, form of the definitions of Granger causality. The idea of improving a prediction is generalized by encoding it into conditional dependence or independence. 

\subsection{From prediction-based definitions\ldots}

Consider a cost function $g:\R^k\longrightarrow \R$ ($k$ is some appropriate dimension), and the associated risk $E[g(e)]$, where $e$ stands for an error term. Let 
a predictor of $x_B(n)$ be defined formally as  $\widehat{x_B}(n+1) = f(x_A^n)$,  where $A$ and $B$ are  subsets of $V$, and $f$ is a function between appropriate spaces, chosen to minimize the risk with $e(n):=x_B(n+1)-\widehat{x_B}(n+1)$. 
Solvability may be granted if $f$ is restricted to an element of a given class of functions, such as the set of linear functions. 
 Let ${\cal F}$ be such a function class. Define:
\begin{eqnarray}
R_{\cal F}\big(B(n+1)\big|A^n\big) = \inf_{f \in {\cal F}} E\big[g\big(x_B(n+1) -  f(x_A^n) \big)\big]
\label{risk:eq}
\end{eqnarray}
$R_{\cal F}\big(B(n+1)\big|A^n\big)$ is therefore the optimal risk when making a one-step-ahead prediction of the multivariate
signal $x_B$ from the past samples of the multivariate signal $x_A$. 
We are now ready to measure the influence of the past of a process on the prediction of another.
To be relatively general  and to prepare comments on the structure of the graph, this can be done for subsets of $V$.
We thus choose  $A$ and $B$ to be two disjoint subsets of $V$, and we define $C:=V\backslash (A\cup B)$ (we use $\backslash$ to mean substraction of a set).  We study causality 
from $x_A$ to $x_B$ by measuring the decrease in the quality of the prediction of $x_B(n)$ when excluding the past of $x_A$.

Let $R_{\cal F}\big(B(n+1)\big|V^n\big)$ be the optimal risk obtained for the prediction of $x_B$ from the past of all of the signals grouped in $x_V$. This risk is compared to $R_{\cal F}\big(B(n+1)\big|(V\backslash A)^n\big)$, where the past of $x_A$ is omitted. Then, for the usual costs functions, we have necessarily:
\begin{eqnarray}
R_{\cal F}\big(B(n+1)\big|V^n\big)  \leq R_{\cal F}\big(B(n+1)\big|(V\backslash A)^n\big)
\end{eqnarray}
A natural first definition for Granger causality is:
\begin{definition} 
$x_A$ Granger  does not cause $x_B$ relative to $V$ if and only if $R_{\cal F}\big(B(n+1)\big|V^n\big)  =  R_{\cal F}\big(B(n+1)\big|(V\backslash A)^n\big)$
\end{definition}
This definition of Granger causality  depends on the cost $g$ chosen as well as on the class ${\cal F}$ of the functions considered.
Usually, a quadratic cost function is chosen, for its simplicity and for its evident physical interpretation (a measure of the power of the error). The choice of the class of functions ${\cal F}$ is crucial. The result of the causality test in definition 1 can change  when the class is changed. Consider the very simple example of
$x_{n+1}=\alpha x_{n} + \beta y^2_{n} +\varepsilon_{n+1}$, where $y_n$ and $\varepsilon_{n}$ are Gaussian independent and identically distributed (i.i.d.) sequences that are  independent of each other. 
The covariance between 
$x_{n+1}$ and $y_n$ is zero, and using the quadratic loss and the class of linear functions, we conclude that $y$ does not Granger cause $x$, because using a linear function of $x_n,y_n$ to predict $x$ would lead to the same minimal risk as using a linear function of $x_n$ only. However, $y_n$ obviously causes $x_n$, but in a nonlinear setting.

The definition is given using the negative of the proposition. If by using the positive way, {\it i.e.,}
$R_{\cal F}\big(B(n+1)\big|V^n\big)   < R_{\cal F}\big(B(n+1)\big|(V\backslash A)^n\big)$, Granger proposes to say that $x_A$ is a {\it prima facie} cause of $x_B$ relative to $V$,  {\it prima facie} can be translated as 'at a first glance'. This is used to insist that if $V$ is enlarged by including other measurements, then the conclusion might be changed. This can be seen as redundant with the mention of the relativity to the observation set $V$, and we therefore do not use this terminology. However, a mention of the relativity to $V$ must be used, as modification of this set can alter the conclusion. A very simple example of this situation is the chain $x_n \rightarrow y_n \rightarrow z_n$, where, for example, $x_n$ is an i.i.d. sequence, $y_{n+1}=x_n + \varepsilon_{n+1}$, $z_{n+1}=y_n + \eta_{n+1}$, $\varepsilon_n, \eta_n$ being independent i.i.d. sequences.  Relative to $V=\{x,z\}$, $x$ causes $z$ if we use the quadratic loss and linear functions of the past samples of $x$ (note here that the predictor $z_{n+1}$ must be a function of not only $x_n$, but also of $x_{n-1}$). However, if we include the past samples of $y$  and $V=\{x,y,z\}$, then the quality of the prediction of $z$ does not deteriorate if we do not use past samples of $x$. Therefore, $x$ does not cause $z$ relative to $V=\{x,y,z\}$. 

The advantage of the prediction-based definition  is that is leads to operational tests. If the quadratic loss is chosen,
working in a parameterized class of functions, such as linear filters or Volterra filters, or even working in reproducing kernel Hilbert spaces, 
allows the implementation of the definition \cite{MariPS08,AmblMRH12,AmblVMR12}. In such cases, the test needed can be evaluated efficiently from the data. From a theoretical point of view, the quadratic loss can be used to find the optimal function in a much wider class of functions: the measurable functions. 
In this class, the optimal function for the quadratic loss is widely known to be the conditional expectation \cite{LehmC98}.  When predicting 
$x_B$ from the whole observation set $V$, the optimal predictor is written as $\widehat{x_B}(n+1) = E\big[x_B(n+1) \big| x_V^n    \big]$.
Likewise, elimination of $A$ from $V$ to study its influence on $B$ leads to the predictor $\widehat{x_B}(n+1) = E\big[x_B(n+1) \big| x_B^n,x_C^n    \big]$, where $V=C\cup A\cup B$. These estimators are of little use, because they are too difficult, or even impossible, to compute. However, they highlight the important of conditional distributions $p(x_B(n+1) \big| x_V^n)$ and $p(x_B(n+1) \big| x_B^n,x_C^n )$ in the problem of testing whether $x_A$ Granger causes $x_B$ relative to $V$ or not.

\subsection{\ldots to a probabilistic definition}

The optimal predictors studied above are equal if the conditional probability distributions 
$p(x_B(n+1) \big| x_V^n)$ and $p(x_B(n+1) \big| x_B^n,x_C^n )$ are equal. These distributions are identical if and only if $x_B(n+1)$ and $x_A^n$ are independent conditionally to $x_B^n,x_C^n $. A natural extension of definition 1 relies on the use of conditional independence. Once again, let $A\cup B \cup C$ be a partition of $V$.
\begin{definition} 
$x_A$  does not Granger cause $x_B$ relative to $V$ if and only if $x_B(n+1) \upmodels  x_A^n  \mid x_B^n,x_C^n, \hspace{.2cm}    \forall n\in \Z$
\end{definition}
This definition means that conditionally to the past of $x_C$, the past of $x_A$ does not bring more information about $x_B(n+1)$ than is  contained in the past of $x_B$. 

Definition 2 is far more general than definition 1. If $x_A$  does not Granger cause $x_B$ relatively to $V$
in the sense of definition 1, it also does not in the sense of definition 2.
Then, definition 2 does not rely on any function class and on any cost function. However, it lacks an inherent operational character: the tools to evaluate conditional independence remain to be defined.
The assessment of conditional independence can be achieved using measures of conditional independence, and some of these measures will be the cornerstone to link directed information theory and Granger causality.

Note also that the concept of causality in this definition is again a relative concept, and that adding or deleting data from the observation set $V$ might modify the conclusions.

\subsection{Instantaneous coupling}
\label{instantcoupl:ssec}

The definitions given so far concern the influence of the past of one process on the present of another one. This is one reason that justifies the use of the term 'causality', when the definitions are actually based on statistical dependence. For an extensive discussion on the differences between causality and  statistical dependence, we refer to \cite{Pear00}.

There is another influence between the processes that is not taken into account by definitions 1 and 2. This influence is referred to as 
'instantaneous causality' \cite{Gran63,Gran80}. However, we will use our preferred term of 'instantaneous coupling', specifically to insist that it is not equivalent to a causal link {\it per se}, but actually a statistical dependence relationship.  The term 'contemporaneous conditional independence' that is used in  \cite{Eich11} could also be chosen. 

Instantaneous coupling measures the common information between $x_A(n+1)$ and $x_B(n+1)$ that is not shared with their past.
A definition of instantaneous coupling might then be that $x_A(n+1)$ and $x_B(n+1)$ are not instantaneously coupled if $x_A(n+1) \upmodels x_B(n+1) \mid  x_A^n,x_B^n, \hspace{.2cm}    \forall n$. This definition makes perfect sense if the observation set is reduced to $A$ and $B$, a situation we refer to as the bivariate case. However, in general, there is also side information $C$, and the definition
 must include this knowledge. However, this presence of side information then leads to two possible definitions of instantaneous coupling.

\begin{definition}
$x_A$ and $x_B$ are not  conditionally instantaneously coupled relative to $V$ if and only if
$x_A(n+1) \upmodels x_B(n+1) \mid  x_A^n,x_B^n,x_C^{n+1}, \hspace{.2cm}    \forall n \in \Z$, where $A\cup B\cup C$ is a partition of $V$.
\end{definition}
The second possibility is the following:
\begin{definition}
$x_A$ and $x_B$ are not  instantaneously coupled relative to $V$  if and only if
$x_A(n+1) \upmodels x_B(n+1) \mid  x_A^n,x_B^n,x_C^{n}, \hspace{.2cm}    \forall n \in \Z$
 \end{definition}
 \vspace{.4cm}

\noindent
Note that definitions 3 and 4 are  symmetrical in $A$ and $B$ (the application of Bayes theorem). 
The difference between definitions 3 and 4 resides in the conditioning on $x_C^{n+1}$  instead of $x_C^{n}$.
 
If the side information up to time $n$ is considered only as in definition 4, the instantaneous dependence or independence is not conditional on the presence of the remaining nodes in $C$. Thus, this coupling is a bivariate instantaneous coupling: it does measure instantaneous dependence (or independence between $A$ and $B$) without considering the possible instantaneous coupling between either $A$ and $C$ or $ B$ and $C$.  Thus, instantaneous coupling found with definition 4 between $A$ and $B$ does not preclude the possibility that the coupling is actually due to couplings between $A$ and $C$ and/or $B$ and $C$. 

Inclusion of all of the information up to time $n+1$ in the conditioning variables allows the dependence or independence to be tested between $x_A(n+1)$ and $x_B(n+1)$ {\em conditionally } to $x_C(n+1)$.

We end up here with the same differences as those between correlation and partial correlation, or dependence and conditional independence for random variables. In graphical modeling, the usual graphs are  based on conditional independence between variables \cite{Whit89,Laur96}. These conditional independence graphs are preferred to independence graphs because of their geometrical properties ( {\it e.g.,} d-separation, \cite{Pear00}), which match the Markov properties possibly present in the multivariate distribution they represent.  From a physical point of view, conditional independence might be preferable, specifically to eliminate 'false' coupling due to third parties.
In this respect, conditional independence is not the panacea, as independent variables can be conditionally dependent. The well-known example is the conditional coupling of independent $x$ and $y$ by their addition. Indeed, even if independent, $x$ and $y$ are conditionally dependent to $z=x+y$. 

\subsection{More on graphs}

Granger causality graphs were defined and studied in \cite{Eich11}. A causality graph is a mixed graph $(V,E_d,E_u)$ that encodes Granger
causality relationships between the components of $x_V$.
The vertex set $V$ stores the indexes of the components of $x_V$. $E_d$ is a set of directed edges beween vertices.
 A directed edge from $a$ to $b$ is equivalent to ``$x_a$ Granger causes $x_b$ relatively to $V$''. $E_u$ is a set of undirected edges. An undirected edge between $x_a$ and $x_b$ 
is equivalent to ``$x_a$ and $x_b$ are (conditionally if def.4 adopted) instantaneously coupled''.
Interestingly, a Granger causality graph may have Markov properties (as in usual graphical models) reflecting a particular (spatial) structure of the joint probability distribution of the whole process $\{x_V^t\} $ \cite{Eich11}.
A taxonomy of Markov properties: local, global, block recursive is studied in \cite{Eich11}, and equivalence between these properties is put forward. More interestingly, these properties are linked with topological properties of the graph. Therefore, structural properties of the graphs are equivalent to a particular factorization of the joint probability of the multivariate process. We will not continue on this subject here, but this must be known since it paves the way to more efficient inference methods for Granger graphical modeling of multivariate processes (see a first step in this direction in \cite{QuinKC11}).

\section{Directed information theory and directional dependence}
\label{infotheory:sec}

Directed information theory  is a recent extension of information theory, even if its roots go back to the 1960's and 1970's and the studies of Marko \cite{Mark73}. The developments began in the late 1990's, after the {\it impetus } given by James Massey in 1990
\cite{Mass90}. The basic theory was then extended by Gerhard Kramer \cite{Kram98}, and then further developed by many authors \cite{Tati00,TatiM09,VenkP07,Kim08,PermKW11} to cite a few. We provide here a short review of the essentials of directed information theory. We will, moreover, adopt a presentation close to the spirit of Granger  causality to highlight the links between Granger causality and information theory. We begin by recalling some basics from information theory. Then, we describe the information-theoretic approach to study directional dependence between stochastic processes, first in the bivariate case, and then, from section \ref{sideinfo:ssec}, for networks, {\it i.e.,} the multivariate case. 

\subsection{Notation and basics}

Let  $H(x_A^n)=-E[\log p(x_A^n )] $ be the entropy of a  random vector $x_A^n$, the density of which is $p$. 
Let the conditional entropy be defined as $ 
H(x_A^n | x_B^n )=-E[\log p(x_A^n | x_B^n )] $.
The mutual information $I(x_A^n ; y_B^n)$  between   $x_A^n$ and $x_B^n$ is defined as \cite{CoveT06}: 
\begin{eqnarray}
I(x_A^n ; x_B^n) &=& H( x_B^n )-H(x_B^n| x_A^n) \nonumber \\ 
&=& D_{KL}\left( p(x_A^n ,x_B^n)\big\| p(x_A^n ) p(x_B^n)\right)
\end{eqnarray}
where $D_{KL}(p||q)= E_p[\log p(x)/q(x)] $ is the Kulback-Leibler divergence.  $D_{KL}(p||q)$ is 0 if and only if $p=q$, and it is positive otherwise.  The mutual information effectively measures independence since it is 0 if and only if  $x_A^n$ and $x_B^n$ are independent random vectors. As   $I(x_A^n ; x_B^n) = I( y_B^n ; x_A^n )$, mutual information  cannot handle directional dependence. 

Let $x_C^n$ be a third time series. It might be a multivariate process that accounts for side information (all of the available observations, but $x_A^n$ and $x_B^n$).  To account for  $x_C^n$, the conditional mutual information is introduced: 
\begin{eqnarray}
 I(x_A^n ; y_B^n |x_C^n )  &= &E\big[ D_{KL}\big( p(x_A^n ,y_B^n| x_C^n)|| p(x_A^n | x_C^n ) p(y_B^m| x_C^n)\big)\big] \\
		&=& D_{KL}\big(p(x_A^n ,y_B^n , x_C^n) || p(x_A^n | x_C^n ) p(y_B^n| x_C^n) p(x_C^n) \big)
\end{eqnarray}
 $ I(x_A^n ; y_B^n | x_C^n ) $ is  zero if and only if  $x_A^n  $ and $y_B^n$ are independent {\em conditionally} to $x_C^n$. 
Stated differently,  conditional mutual information measures the divergence between the actual observations and those which would be observed under the Markov  assumption $(x \rightarrow  z \rightarrow y)$. Arrows can be misleading here, as by reversibility of Markov chains, the equality above holds also for $(y \rightarrow  z \rightarrow x)$. This emphasizes how mutual information cannot provide answers to the information flow directivity problem.

\subsection{Directional dependence between stochastic processes; causal conditioning}

The dependence between the components of  the stochastic process $x_V$  is encoded in the full generality by the joint probability distributions $p(x_V^n)$. If $V$ is partitioned into subsets $A,B,C$, studying dependencies between $A$ and $B$  then requires that $p(x_V^n)$ is factorized into terms where $x_A$ and $x_B$ appear. For example, as
$p(x_V^n)=p(x_A^n,x_B^n,x_C^n)$, we can factorize the probability distribution as $p(x_B^n | x_A^n, x_C^n) p (x_A^n,x_C^n)$, which appears to emphasize a link from $A$ to $B$. Two problems appear, however: first, the presence of $C$ perturbs the analysis (more than this, $A$ and $C$ have a symmetrical role here); secondly, the factorization does not take into account the arrow of time, as the conditioning is considered over the whole observations up to time $n$. 

Marginalizing $x_C$ out makes it possible to work directly on $p(x_A^n,x_B^n)$.  However, this  eliminates all of the dependence between $A$ and $B$ that might exist {\it via}  $C$, and therefore  this might lead to an incorrect assessment of the dependence. As for Granger causality, this means that dependence analysis is relative to the observation set. Restricting the study to $A$ and $B$ is what we referred to as the bivariate case, and this allows the basic ideas to be studied. We will therefore present directed information first in the bivariate case, and then turn to the full multivariate case. 

The second problem is at the root of the measure of  directional dependence between stochastic processes. Assuming that $x_A(n)$ and $x_B(n)$ are linked by some physical ({\it e.g.,} biological, economical) system, it is natural to postulate that their dependence is constrained by causality: if $A\rightarrow B$, then an event occurring at some time in $A$ will influence  $B$  later on.  Let us come back to the simple factorization above for the bivariate case. We have $p(x_A^n,x_B^n)= p(x_B^n | x_A^n) p (x_A^n)$, and furthermore\footnote{We implicitly choose 1 here as the initial time.}:
\begin{eqnarray}
p(x_B^n | x_A^n) &=& \prod_{i=1}^n p\big(x_B(i)  \big| x_B^{i-1}, x_A^n\big) 
\label{distcond:eq}
\end{eqnarray}
where for  $i=1$, the first term is  $p(x_B(1)|x_A(1))$. 
The conditional distribution quantifies a directional dependence from $A$ to $B$, but it lacks the causality property mentioned above, as $p\big(x_B(i)  \big| x_B^{i-1}, x_A^n\big)$ quantifies the influence of the whole observation $x_A^n$ (past and future of $i$) on the present $x_B(i)$ knowing its past $x_B^{i-1}$. The causality principle would require the restriction of the {\it prior} time $i$ to the past of $A$ only. 
Kramer defined 'causal conditioning' precisely in this sense \cite{Kram98}. Modifying Eq. (\ref{distcond:eq})
accordingly, we end up we the definition of the causal conditional probability distribution:
\begin{eqnarray}
p(x_B^n  \| x_A^n) &:=& \prod_{i=1}^n p\big(x_B(i)  \big| x_B^{i-1}, x_A^i\big)
\label{distcausalcond:eq}
\end{eqnarray}
Remarkably  this provides an alternative factorization of the joint probability.
As noted by Massey \cite{Mass90},     $p(x_A^n,y_B^n)$ can then be factorized as\footnote{$x_B^{n-1}$ stands for the delayed collections of samples of $x_B$. If the time origin is finite, 0 or 1, the first element of the list $x_B^{n-1}$ 
should be understood as a wild card $\emptyset$ which does not influence the conditioning.}: 
\begin{eqnarray}
p(x_A^n,x_B^n )  & =& p(x_B^n \| x_A^n)  p (x_A^n\|x_B^{n-1} )
\label{factorisation:eq}
\end{eqnarray}
Assuming that $x_A$ is the input of a 
 system that creates $x_B$,  $p (x_A^n\|x_B^{n-1} )=\prod_i p(x_A(i) | x_A^{i-1},x_B^{i-1}) $ characterizes   the  feedback in the system: each of the factors controls the probability of the input $x_A$ at time $i$ conditionally to its past and to the past values of the output $x_B$.
Likewise, the term $p(x_B^n \| x_A^n) =\prod_i  p(x_B(i) | x_B^{i-1},x_A^{i}) $ characterizes the direct (or feedforward) link in the system.

Several interesting simple cases occur:
\begin{itemize}
\item  In the absence of feedback in the link from $A$ to $B$, there is the following:
\begin{eqnarray}
p(x_A(i) \big| x_A^{i-1}, x_B^{i-1} ) = p(x_A(i) \big| x_A^{i-1}),   \mbox{ } \forall i\geq 2
\end{eqnarray}
or equivalently, in terms of entropies, 
\begin{eqnarray}
H(x_A(i) \big| x_A^{i-1}, x_B^{i-1} ) = H(x_A(i) \big| x_A^{i-1}) ,  \mbox{ } \forall i\geq 2
\end{eqnarray}
and as  a consequence: 
\begin{eqnarray}
p (x_A^n\|x_B^{n-1} )= p(x_A^n)
\end{eqnarray}
\item Likewise, if there is only a feedback term, then $p(x_B(i) | x_B^{i-1},x_A^{i})=p(x_B(i) | x_B^{i-1})$ and then:
\begin{eqnarray}
p(x_B^n \| x_A^n) = p(x_B^n)
\end{eqnarray}
\item  If the link is  memoryless, {\it i.e.,} the output $x_B$ does not depend on the past, then:
\begin{eqnarray}
 p(x_B(i) | x_A^{i},y_B^{i-1})=p(x_B(i) \big| x_A(i))  \mbox{ } \forall i\geq 1
\end{eqnarray}
\end{itemize}
%For future need, let $D$ be the unit delay operator, such that $Dx_B(n)=x_B(n-1)$. We define $Dx_B^n= (\emptyset, x_B(1), x_B(2),\ldots, x_B(n-1)) $ for finite length sequences, in order to  deal with edge effects while maintaining constant dimension for the studied time series. The term $\emptyset$  indicates a wild card which plays no influence on conditioning, and makes sense as $x_B(0)$ is not assumed observed. 

These results allow the question of whether $x_A$ influences $x_B$ to be addressed.
If it does, then the joint distribution has the factorization of Eq. (\ref{factorisation:eq}). However, if $x_A$ does not influence $x_B$, 
then $p(x_B^n \| x_A^n) = p(x_B^n)$, and the factorization of the joint probability distribution simplifies to $p (x_A^n\|x_B^{n-1} )p(x_B^n)$.
Kullback divergence between the probability distributions for each case generalizes the definition of mutual information to the directional mutual information:
\begin{eqnarray}
I(x_A^n \rightarrow x_B^n )  = D_{KL} \left(  p(x_A^n,x_B^n) \big\| p(x_A^n \|x_B^{n-1} ) p(x_B^n)\right)  \label{infodirKL:eq}
\end{eqnarray}
This quantity measures the loss of information when it is incorrectly assumed that $x_A$ does not influence $x_B$. This was called {\em directed information} by Massey \cite{Mass90}. 
Expanding the Kullback divergence allows different forms for the directed information to be obtained:
\begin{eqnarray}
I(x_A^n \rightarrow x_B^n )  &=& \sum_{i=1}^n I\big(x_A^i ; x_B(i) \big| x_B^{i-1}\big) \\
&=&H\big(x_B^n ) - H\big(x_B^n \big\|  x_A^n\big)
\end{eqnarray}
where we define the `causal conditional entropy':
\begin{eqnarray}
H\big(x_B^n \big\| x_A^n\big) &=&-E\big[ \log p\big(x_B^n \big\|  x_A^n\big)   \big] \\
&=& \sum_{i=1}^n H\big( x_B(i) \big| x_B^{i-1}, x_A^i \big) 
\end{eqnarray}
Note that causal conditioning might involve more than one process. This leads to the defining of the causal conditional directed information as: 
\begin{eqnarray}
I(x_A^n \rightarrow x_B^n  \| x_C^n) &:=& H\big(x_B^n \big\|  x_C^n\big)  - H\big(x_B^n \big\|  x_A^n,x_C^n\big) \nonumber \\
&=& \sum_{i=1}^n I\big(x_A^i ; x_B(i) \big| x_B^{i-1}, x_C^i\big)
\label{causaldi1:eq}
\end{eqnarray}

The basic properties of the directed information were studied by Massey and Kramer \cite{Mass90,MassM05,Kram98}, and some are recalled below.  As a Kullback divergence, the directed information is always positive or zero. 
Then, simple algebraic manipulation allows the decomposition to be obtained:
\begin{eqnarray}
I(x_A^n \rightarrow x_B^n )+   I( x_B^{n-1} \rightarrow x_A^n ) 
&=& I( x_A^n ; x_B^n )  \label{decompdi:eq}
\end{eqnarray}
Eq. (\ref{decompdi:eq}) is fundamental, as it shows how mutual information splits into the sum of a feedforward information flow $I(x_A^n \rightarrow x_B^n )$ and a feedback information flow $I( x_B^{n-1} \rightarrow x_A^n )$. 
In the absence of feedback,  $p(x_A^n \| x_B^{n-1})  = p(x_A^n)$ and $I(x_A^n;x_B^n )=I(x_A^n \rightarrow x_B^n )$.  Eq.  (\ref{decompdi:eq}) allows the conclusion that the mutual information is always greater than the directed information, as  $ I( x_B^{n-1} \rightarrow x_A^n )$ is always positive or zero (as directed information). It is zero if and only if: 
\begin{eqnarray}
I(x_A(i) ; x_B^{i-1} \big| x_A^{i-1} ) = 0 \mbox{ } \forall i=2,\ldots,n
\label{infonofeedback:eq}
\end{eqnarray}
or equivalently:
\begin{eqnarray}
H(x_A(i) \big| x_A^{i-1},x_B^{i-1}  ) = H(x_A(i) \big| x_A^{i-1} ) \mbox{ } \forall i=2,\ldots,n
\label{entropienofeedback:eq}
\end{eqnarray}
This situation  corresponds to the absence of feedback in the link $A\rightarrow B$, whence the fundamental result that the directed information and the mutual information are equal if the channel is free of feedback. 
This result implies  that mutual information over-estimates  the directed information between two processes in the presence of feedback. This was thoroughly studied  in  \cite{Kram98,Tati00,VenkP07,TatiM09}, in a communication-theoretic framework.

The decomposition of Eq. (\ref{decompdi:eq}) is surprising, as it shows that the mutual information is not the sum 
of the directed information flowing in both directions. Instead, the following decomposition holds:
\begin{eqnarray}
I(x_A^n \rightarrow x_B^n ) + I(x_B^n \rightarrow x_A^n ) &=&  I(x_A^n ; x_B^n ) + I(x_A^n \rightarrow x_B^n \| x_A^{n-1} )
\label{sums2di:eq}
\end{eqnarray}
where: 
\begin{eqnarray}
 I(x_A^n \rightarrow x_B^n || x_A^{n-1} )  &=&  \sum_i I(x_A^{i};x_B(i)|x_B^{i-1},x_A^{i-1}) \nonumber \\
 & =& \sum_i I(x_A(i);x_B(i)|x_B^{i-1},x_A^{i-1}) 
\end{eqnarray}
 This demonstrates that 
$I(x_A^n \rightarrow x_B^n ) + I(x_B^n \rightarrow x_A^n ) $ is
symmetrical, but is in general not equal to the mutual information, 
except if and only if  $ I(x_A(i);x_B(i)|x_B^{i-1},x_A^{i-1})=0, \forall i=1,\dots,n$.  As the term in the sum is the mutual information between the present samples of the two processes conditioned on their joint past values, this measure is a measure of instantaneous dependence. It is indeed symmetrical in $A$ and $B$. 
The term 
$I(x_A^n \rightarrow x_B^n || x_A^{n-1} )  = I(x_B^n \rightarrow x_A^n || x_B^{n-1} ) $ will thus be named the {\em instantaneous information exchange}  between   $x_A$ and $x_B$, and will hereafter be denoted as $I(x_A^n \leftrightarrow x_B^n)$. Like directed information, conditional forms of the instantaneous information exchange can be defined, as for example:
\begin{eqnarray}
I(x_A^n \leftrightarrow x_B^n \| x_C^n) := I(x_A^n \rightarrow x_B^n || x_A^{n-1}, x_C^n ) 
\end{eqnarray}
which quantifies an instantaneous information exchange between $A$ and $B$ causally conditionally to $C$.
%%%%%%%%%%%%%%% 

\subsection{Directed information rates}

Entropy and mutual information  in general  increase linearly with 
the length $n$ of the recorded time series.  
 Shannon's information rate for stochastic processes compensates for the linear growth by considering  $A_\infty(x)= \lim_{n\rightarrow+\infty} A(x^n) / n $ ( if the limit exists), where $A(x^n)$
denotes any information measure on the sample $x^n$ of length $n$. 
 
For the important class of stationary processes (see {\it e.g.,} \cite{CoveT06}),  the entropy rate turns out to be the limit of the conditional entropy: 
\begin{eqnarray}
\lim_{n\rightarrow+\infty}\frac{1}{n} H(x_A^n)  =  \lim_{n\rightarrow+\infty} H(x_A(n)| x_A^{n-1})
\end{eqnarray}
 Kramer generalized this result for causal conditional entropies \cite{Kram98}, thus defining 
 the directed information rate for stationary processes  as:
 \begin{eqnarray}
I_\infty(x_A\rightarrow x_B) &=& \lim_{n\rightarrow+\infty} \frac{1}{n}\sum_{i=1}^{n} I( x_A^{i} ; x_B(i) |x_B^{i-1}) \nonumber \\
&=& \lim_{n\rightarrow+\infty} I( x_A^n ; x_B(n) |x_B^{n-1})
\label{eq:rateDirInfo}
\end{eqnarray}
This result holds also for the instantaneous information exchange rate. 
 Note that the proof of the result relies on the positivity of the entropy for discrete valued stochastic processes. 
For continously valued processes, for which the entropy can be negative, the proof is more involved and requires the methods developed  in \cite{Pins64,GrayK80,Gray90}, and see also \cite{TatiM09}. 

%%%%%%%%

\subsection{Transfer entropy and instantaneous information exchange}

As introduced by Schreiber in  \cite{Schr00,KaisS02}, {\em transfer entropy} evaluates the deviation of the observed data from a model, assuming the following  joint Markov property:
\begin{eqnarray}
p(x_B(n) | x_{B  \,  n-k+1}^{\mbox{ } \,\,\,\, n-1}, x_{A \,  n-l+1}^{\mbox{ } \,\,\,\, n-1} ) = p(x_B(n) | x_{B  \,  n-k+1}^{\mbox{ } \,\,\,\, n-1})
\label{eq:jointMarkov}
\end{eqnarray}
This leads to the following definition:
\begin{eqnarray}
T(x_{A \,  n-l+1}^{\mbox{ } \,\,\,\, n-1}\rightarrow x_{B  \,  n-k+1}^{\mbox{ } \,\,\,\, n})  = E\left[  \log \frac{p(x_B(n) | x_{B  \,  n-k+1}^{\mbox{ } \,\,\,\, n-1}, x_{A \,  n-l+1}^{\mbox{ } \,\,\,\, n-1} ) }{p(x_B(n) | x_{B  \,  n-k+1}^{\mbox{ } \,\,\,\, n-1})} \right]
\end{eqnarray}
Then $T(x_{A \,  n-l+1}^{\mbox{ } \,\,\,\, n-1} \rightarrow x_{B  \,  n-k+1}^{\mbox{ } \,\,\,\, n} ) =0 $ if and only if Eq. (\ref{eq:jointMarkov}) is satisfied.
Although in the original definition, the past of $x$ in the conditioning might  begin at a different time $m \not=n$,  for practical reasons $m=n$ is considered. Actually, no {\it a priori } information is available about possible delays, and setting $m=n$ allows  the transfer entropy to be compared with the directed information. 

 By expressing  the transfer entropy as a difference of conditional entropies, we get: 
\begin{eqnarray}
T(x_{A \, n-l+1}^{\mbox{ } \,\,\,\, n-1} \rightarrow  x_{B  \, n-k+1}^{ \mbox{ } \,\,\,\, n} )  &= & H(x_B(n) | x_{B  \, n-k+1}^{ \mbox{ } \,\,\,\, n-1}) - H(x_B(n) |x_{B  \, n-k+1}^{ \mbox{ } \,\,\,\, n-1}, x_{A \, n-l+1}^{ \mbox{ } \,\,\,\, n-1} ) \nonumber\\
&=& I(x_{A \, n-l+1}^{\mbox{ } \,\,\,\, n-1}  ; x_B(n)  |x_{B  \, n-k+1}^{ \mbox{ } \,\,\,\, n-1})
\end{eqnarray}
For $l=n=k$ and choosing 1 as the time origin,  the identity
 $I(x,y ; z|w)=I(x;z|w)+I(y;z|x,w)$  leads to:
 \begin{eqnarray}
 I( x_A^n ; x_B(n) |x_B^{n-1}) &=& I( x_A^{n-1} ; x_B(n) |x_B^{n-1}) + I(x_A(n) ; x_B(n) |x_A^{n-1} , x_B^{n-1}) \nonumber \\
	 &= &T(x_A^{n-1} \rightarrow  x_B^n ) +  I(x_A(n) ; x_B(n)|x_A^{n-1} , x_B^{n-1})
	 \label{eq:SchreiberT}
\end{eqnarray}
For stationary processes, letting $n\rightarrow \infty$ and  provided the limits exist, for the rates, we obtain:
 \begin{eqnarray}
I_\infty( x_A \rightarrow x_B) = T_\infty(x_A \rightarrow x_B ) +  I_\infty(x_A \leftrightarrow x_B )
\end{eqnarray}
Transfer entropy is the part of the directed information that measures the influence of the past of $x_A$ on the present of $x_B$. However it does not take into account the possible instantaneous  dependence of one time series on another, which is handled by directed information. 

Moreover,  as defined by Schreiber in \cite{Schr00,KaisS02}, only $ I( x_A^{i-1} ; x_B(i) |x_B^{i-1})$ is considered in $T$, instead of  its sum over $i$  in the directed information. Thus stationarity is implicitly assumed and the transfer entropy has  the same 
meaning as  a rate. A sum over delays was considered by Palu\v{s} as a means of reducing errors when estimating the measure \cite{PaluV07}.
Summing 
over $n$ in Eq. (\ref{eq:SchreiberT}),  the following  decomposition of the directed information is obtained:
\begin{eqnarray}
I(x_A^n \rightarrow  x_B^n ) = I( x_A^{n-1} \rightarrow  x_B^n ) + I(x_A^n \leftrightarrow  x_B^n )
\label{dirinfodecomp:eq}
\end{eqnarray}
Eq. (\ref{dirinfodecomp:eq}) establishes that the influence of one process on another can be decomposed into two terms that account for the past and for the instantaneous contributions. Moreover, this explains the presence of the term $I(x_A^n \leftrightarrow  x_B^n )$ in the r.h.s. of Eq. (\ref{sums2di:eq}): Instantaneous information exchange is counted twice in the l.h.s. terms 
$I(x_A^n \rightarrow x_B^n ) + I(x_B^n \rightarrow x_A^n ) $, but only once in the mutual information $ I(x_A^n ; x_B^n )$. This allows Eq. (\ref{sums2di:eq}) to be written in a slightly different form, as:
\begin{eqnarray}
I(x_A^{n-1} \rightarrow x_B^n ) + I(x_B^{n-1} \rightarrow x_A^n ) +I(x_A^n \leftrightarrow  x_B^n ) &=&  I(x_A^n ; x_B^n ) \label{sums2diinst:eq}
\end{eqnarray}
which is very appealing, as it shows how dependence as measured by mutual information  decomposes as the sum of the measures of directional dependences and the measure of instantaneous coupling. 

%%%%%
%toto
\subsection{Accounting for side information}
\label{sideinfo:ssec}

The preceding  developments  aimed at the proposing of definitions of the information flow between $x_A$ and $x_B$;  however, whenever $A$ and $B$ are connected to other parts of the network, the flow of information between $A$ and $B$ might be mediated by other members of the network.  Time series observed on nodes other than $A$ and $B$ are hereafter referred to as side information. The available side information at time $n$ is denoted as $x_C^n$, with $A,B,C$ forming a partition of $V$. Then, depending on the type of conditioning (usual or causal) two approaches are possible. Usual conditioning considers directed information from $A$ to $B$ that is conditioned on the whole observation $x_C^n$. However, this leads to the consideration of causal flows from $A$ to $B$ that possibly include a flow that goes from $A$ to $B$ {\it via } $C$ in the future! Thus, an alternate definition for conditioning is required. This is given by the definition of Eq. (\ref{causaldi1:eq}) of the causal conditional directed information:
\begin{eqnarray}
I(x_A^n \rightarrow x_B^n  \| x_C^n) &:=& H\big(x_B^n \big\|  x_C^n\big)  - H\big(x_B^n \big\|  x_A^n,x_C^n\big) \nonumber \\
&=& \sum_{i=1}^n I\big(x_A^i ; x_B(i) \big| x_B^{i-1}, x_C^i\big)
\end{eqnarray}

Does the causal conditional directed information decompose as the sum of a causal conditional transfer entropy and a causal conditional instantaneous information exchange, as it does in the bivariate case? 
Applying twice the chain rule for conditional mutual information, we obtain:
\begin{eqnarray}
I(x_A^n \rightarrow x_B^n \big\| x_C^{n} ) = I(x_A^{n-1} \rightarrow x_B^n \big\| x_C^{n-1} ) +
I(x_A^n \leftrightarrow x_B^n \big\| x_C^{n} ) + \Delta I (x_C^n \leftrightarrow x_B^n  ) 
\label{decomp:eq}
\end{eqnarray}
In this equation, $ I(x_A^{n-1} \rightarrow x_B^n \big\| x_C^{n-1} )$ is termed the 'causal conditional transfer entropy'. This measures the flow of information from $A$ to $B$ by taking into account a possible route {\it via }  $C$. If the flow of information from $A$ to $B$ is entirely
relayed by $C$, the 'causal conditional  transfer entropy' is zero.  In this situation, the usual transfer entropy is not zero, indicating the existence of a flow from $A$ to $B$. Conditioning on $C$ allows the examination of whether the route goes through $C$.
The term:
\begin{eqnarray}
I(x_A^n \leftrightarrow x_B^n \big\|   x_C^{n} ) &:=& I(x_A^n \rightarrow x_B^n \big\| x_A^{n-1}, x_C^{n} )\\
&=& \sum_{i=1}^n I(x_A(i);x_B(i)|x_B^{i-1},x_A^{i-1}, x_C^{i})
\end{eqnarray}
is the 'causal conditional information exchange'. This  measures the conditional instantaneous coupling between $A$ and $B$.  
The term $ \Delta I (x_C^n \leftrightarrow x_B^n  )$ emphasizes the difference between the bivariate and the multivariate cases. This extra term  measures an instantaneous coupling and is defined by:
\begin{eqnarray}
\Delta I (x_C^n \leftrightarrow x_B^n  ) = I (x_C^n \leftrightarrow x_B^n \big\| x_A^{n-1} ) -
 I (x_C^n \leftrightarrow x_B^n  )
\end{eqnarray}
An alternate decomposition to Eq. (\ref{decomp:eq}) is:
\begin{eqnarray}
I(x_A^n \rightarrow x_B^n \big\| x_C^{n} ) = I(x_A^{n-1} \rightarrow x_B^n \big\| x_C^{n} ) +
I(x_A^n \leftrightarrow x_B^n \big\| x_C^{n} ) 
\end{eqnarray}
which  emphasizes that the extra term comes from: 
\begin{eqnarray}
 I(x_A^{n-1} \rightarrow x_B^n \big\| x_C^{n} ) = I(x_A^{n-1} \rightarrow x_B^n \big\| x_C^{n-1} )
  +\Delta I (x_C^n \leftrightarrow x_B^n  ) 
 \end{eqnarray}
This demonstrates that the definition of the conditional transfer entropy requires  conditioning on the past of $C$. If not,  the extra term appears  and accounts for instantaneous information exchanges between $C$ and $B$, due to the addition of the term $x_C(i)$ in the conditioning. 
 This extra term highlights the difference between the two different natures  of instantaneous coupling. The first  term, 
 \begin{eqnarray}
I (x_C^n \leftrightarrow x_B^n \big\| x_A^{n-1} ) = \sum_i I( x_C(i) ; x_B(i) \big| x_A^{i-1} , x_B^{i-1}, x_C^{i-1} )
\end{eqnarray}
 describes the  intrinsic coupling in the sense that it does not depend on parties other than $C$ and $B$. The second coupling  term, 
 \begin{eqnarray*}
I (x_C^n \leftrightarrow x_B^n  ) = \sum_i I(x_C(i); x_B(i) \big| x_B^{i-1}, x_C^{i-1} )
\end{eqnarray*}
 is  relative to the  extrinsic coupling, as it measures the instantaneous coupling at time $i$ that is created by variables other than $B$ and $C$.

As discussed in section \ref{instantcoupl:ssec}, the second definition for instantaneous coupling considers conditioning  on the past of the side information {\em only}. Causally conditioning on $x_C^{n-1}$ does not modify the results of the bivariate case. In particular, we still get the elegant decomposition:
\begin{eqnarray}
I(x_A^n \rightarrow x_B^n \big\| x_C^{n-1} ) = I(x_A^{n-1} \rightarrow x_B^n \big\| x_C^{n-1} ) +
I(x_A^n \leftrightarrow x_B^n \big\| x_C^{n-1} ) 
\end{eqnarray}
and therefore, the decomposition of Eq. (\ref{sums2diinst:eq}) is generalized to:
\begin{eqnarray}
I(x_A^{n-1} \rightarrow x_B^n \big\| x_C^{n-1} ) + I(x_B^{n-1} \rightarrow x_A^n \big\| x_C^{n-1} ) +I(x_A^n \leftrightarrow  x_B^n\big\| x_C^{n-1}  ) &=&  I(x_A^n ; x_B^n \big\| x_C^{n-1} ) 
\label{sums2diinstcond:eq}
\end{eqnarray}
where:
\begin{eqnarray}
 I(x_A^n ; x_B^n \big\| x_C^{n-1} ) = \sum_i I\big(x_A^n; x_B(i) \big| x_B^{i-1},x_C^{i-1} \big)
\end{eqnarray}
is the causally conditioned mutual information.

Finally, let us consider that for jointly stationary times series,  the causal directed information rate is defined similarly to the bivariate case, as: 
\begin{eqnarray}
I_\infty(x_A\rightarrow x_B \big\|  x_C) &=& \lim_{n\rightarrow+\infty} \frac{1}{n}\sum_{i=1}^{n} I\big( x_A^{i} ; x_B(i) |x_B^{i-1}, x_C^{i}\big) \\
&=& \lim_{n\rightarrow+\infty} I\big( x_A^n ; x_B(n) \big| x_B^{n-1} , x_C^{n}\big)
\end{eqnarray}

In this section we have emphasized on Kramer's causal conditioning, both for the definition of directed information and  for taking into account  side information.  We have also shown that Schreiber's transfer entropy is that part of the directed information that is dedicated to the strict sense of causal information flow (not accounting for simultaneous coupling). The next section more explicitely revisits the links between Granger causality and directed information theory.

\section{Inferring Granger causality and instantaneous coupling}
\label{links:sec}

Granger causality in its probabilistic form is not operational. In practical situations,
for assessing Granger causality between time series, we cannot use the definition directly. We have to define  dedicated tools to assess the conditional independence. We use this inference framework to show
the links between information theory and Granger causality. We begin by re-expressing Granger causality definitions in terms of some measures that arise from directed information theory. Therefore, in an inference problem, these measures can be used as tools for inference. However, we show in the following sections that these measures naturally emerge from the more usual statistical inference strategies.
In the following, and as above, we use the same partitioning of $V$ into the union of disjoint subsets of $A$, $B$ and $C$.

\subsection{Information-theoretic measures and Granger causality}

As anticipated in the presentation of directed information, there are profound links between Granger causality and directed information measures. Granger causality relies on  conditional independence, and it can also be defined using measures of conditional independence. Information-theoretic measures appear as natural candidates. 
% Furthermore, it turns out that the measures are precisely the measures that appear in the decompositions of directed information shown in the preceding section.
Recall that two random elements are independent if and only if their mutual information is zero. Moreover, two random elements are independent conditionally to a third one if and only if the  conditional mutual  information is zero. 
We can  reconsider definitions 2, 3 and 4 and recast them in term of information-theoretic measures.

Definition 2 stated that
$x_A$ does not Granger cause $x_B$ relative to $V$ if and only if $x_B(n+1) \upmodels  x_A^n  \mid x_B^n,x_C^n, \hspace{.2cm}    \forall n\geq 1$. This can be alternatively rephrased into:

\begin{definition} 
$x_A$  does not Granger cause $x_B$ relative to $V$ if and only if $I( x_A^{n-1} \rightarrow x_B^n \| x_C^{n-1} )=0 \hspace{.2cm}    \forall n\geq 1$
\end{definition}
since $x_B(i) \upmodels  x_A^i  \mid x_A^{i-1},x_C^{i-1}, \hspace{.2cm}    \forall  1\leq i \leq n$ is equivalent to 
$I(x_B(i);  x_A^i  \mid x_A^{i-1},x_C^{i-1}) =0  \hspace{.2cm}   \forall  1\leq i \leq n$.

Otherwise stated, the transfer entropy from $A$ to $B$ causally conditioned on  $C$ is zero if and only if $A$ does not Granger cause $B$ relative to $V$. 
This shows that causal conditional transfer entropy can be used to assess Granger causality.

Likewise, we can give alternative definitions of instantaneous coupling.

\begin{definition}
$x_A$ and $x_B$ are not  conditionally instantaneously coupled relative to $V$ if and only if 
$I(x_A^n \leftrightarrow x_B^n \big\| x_C^{n} )     \forall n \geq 1$,
\end{definition}
or if and only if the instantaneous information exchange causally conditioned on $C$ is zero.
The second possible definition of instantaneous coupling is equivalent to:
\begin{definition}
$x_A$ and $x_B$ are not  instantaneously coupled relative to $V$  if and only if
$I(x_A^n \leftrightarrow x_B^n \big\| x_C^{n-1} )     \forall n \geq 1$,
 \end{definition}
or if and only if the instantaneous information exchange causally conditioned on the past of  $C$ is zero.

Note that in the bivariate case  only (when $C$ is not taken into account), the directed information 
$I(x_A^n \rightarrow x_B^n)$ summarizes both the Granger causality and the coupling, as it decomposes as the sum of the transfer entropy $I(x_A^{n-1}\rightarrow x_B^n)$ and the instantaneous information exchange $I(x_A^{n-1}\leftrightarrow x_B^n)$.

\subsection{Granger causality inference}

We consider the practical problem of inferring the graph of dependence between the components of a multivariate process. 
Let us assume that we have measured a multivariate process $x_V(n)$ for $n\leq T$. We want to study the dependence between each pair of components (Granger causality and instantaneous coupling between any pair of components relative to $V$).

We can use the result of the preceding section to evaluate the directed information measures on the data. When studying the influence from any subset  $A$ to any subset $B$, if the measures  are zero, then there is no causality (or no coupling); if they are strictly positive, then $A$ Granger causes $B$ relative to $V$ (or $A$ and $B$ are coupled relative to $V$). This point of view has been adopted in many of the studies that we have already referred to ({\it e.g. } \cite{KaisS02,HlavSPVB07,Palu07,QuinCKH11,ViceWLP11}), and it relies on estimating the measures from the data. 
We will not review the estimation problem here.

However, it is interesting  to examine more traditional frameworks  for testing Granger causality, and to examine how directed information theory naturally emerges from these frameworks. To begin with, we show how the measures defined emerge from a binary hypothesis-testing view of Granger causality inference. We then turn to prediction and model-based approaches. We will 
review how  Geweke's measures of Granger causality in the Gaussian case are equivalent to directed information measures. We will then 
present a more general case adopted by \cite{GourMR87,RissW87,KimB10,KimPGB11,QuinCKH11} and based on a model  of the data. 

\subsubsection{Directed information emerges from a hypotheses-testing framework}
 \label{binaryhyp:ssec}
 
In the inference problem, we want to determine whether or not $x_A$ Granger causes  (is coupled with) or not $x_B$ relative to $V$.
This can be formulated as a binary hypothesis testing problem. For  inferring dependencies between $A$ and $B$ relative to $V$, we can state the problem as follows.

Assume we observe $x_V(n), \forall n \leq T$. Then, we want to test: '$x_A$ does not Granger cause $x_B$', against '$x_A$ causes $x_B$'; and 
'$x_A$ and $x_B$ are instantaneously  coupled' against `$x_A$ are $x_B$ not instantaneously  coupled'. We will refer to the first test as the Granger causality test, and to the second one, as the instantaneous coupling test. 

In the bivariate case, for which the Granger causality test indicates:
\begin{eqnarray}
\left\{\begin{array}{lclcl}
H_0    & : & p_0(x_B(i)\mid x_A^{i-1}, x_B^{i-1} )  &=& p( x_B(i)\mid x_B^{i-1} )  , \forall i\leq T \\
H_1     &:  & p_1(x_B(i)\mid x_A^{i-1}, x_B^{i-1} ) &=& p( x_B(i)\mid x_A^{i-1}, x_B^{i-1} ),  \forall i\leq T 
 \end{array}\right.
 \label{test1:eq}
\end{eqnarray}
this leads to the testing of  different functional forms of the conditional densities of $x_B(i)$ given the past of $x_A$.  
The likelihood  of the observation under $H_1$ is the full joint probability $p(x_A^T,x_B^T)= p(x_A^T\| x_B^T) p(x_B^T \| x_A^{T-1} )$. Under $H_0$ we have $p(x_B^T \| x_A^{T-1} )=p(x_B^T)$ and the
likelihood reduces to $p(x_A^T\| x_B^T) p(x_B^T \| x_A^{T-1} )=p(x_A^T\| x_B^T) p(x_B^T)$. The log likelihood {\it ratio } for the test is:
\begin{eqnarray}
l(x_A^T,x_B^T) &:=& \log \frac{p(x_A^T, x_B^T \mid H_1)}{p(x_A^T, x_B^T \mid H_0)}  = \log  \frac{p(x_B^T \| x_A^{T-1} )}{p(x_B^T)}\\
&=& \sum_{i=1}^T \log \frac{p(x_B(i) \mid x_A^{i-1},x_B^{i-1})}{p(x_B(i) \mid x_B^{i-1})}
\end{eqnarray}
For example, in the case where the multivariate process is a positive Harris recurrent Markov chain \cite{MeynT09},  the law of large numbers applies and we have under hypothesis $H_1$:
\begin{eqnarray}
\frac{1}{T}l(x_A^T,x_B^T)   \xrightarrow{T\rightarrow +\infty} T_\infty( x_A \rightarrow x_B)   \mbox{ a.s.}
\end{eqnarray}
where $T_\infty( x_A \rightarrow x_B)$ is the transfer entropy rate. Thus from a practical point of view, as the amount of data increases,
we expect the log likelihood {\it ratio} to be close to the transfer entropy rate (under $H_1$). Turning the point of view, this can justify the use of an estimated transfer entropy to assess Granger causality. Under $H_0$, $\frac{1}{T}l(x_A^T,x_B^T)$ converges to  $\lim_{T\rightarrow +\infty}(1/T) D_{KL}\big(p(x_A^T\| x_B^T) p(x_B^T)   \big\|  p(x_A^T\| x_B^T) p(x_B^T \| x_A^{T-1})   \big)$,
which can be termed  `the Lautum transfer entropy rate' that extends the `Lautum directed information'  defined  in \cite{PermKW11}. Directed information can be viewed as a measure of the loss of information when assuming  $x_A$ does not causally influence $x_B$ when it actually does. Likewise,  `Lautum directed information' measures the loss of information when assuming  $x_A$ does  causally influence $x_B$, when actually it does not. 

For testing instantaneous coupling, we will use the following:
\begin{eqnarray}
\left\{\begin{array}{lclcl}
H_0    & : & p_0(x_A(i),x_B(i)\mid x_A^{i-1}, x_B^{i-1} )  &=& p( x_A(i)\mid x_A^{i-1},x_B^{i-1} ) p( x_B(i)\mid x_A^{i-1},x_B^{i-1} )  , \forall i\leq T \\
H_1     &:  & p_1(x_A(i),x_B(i)\mid x_A^{i-1}, x_B^{i-1} ) &=& p( x_A(i),x_B(i)\mid x_A^{i-1}, x_B^{i-1} ),  \forall i\leq T 
 \end{array}\right.
 \label{test2:eq}
\end{eqnarray}
where under $H_0$, there is no coupling. Then, under $H_1$ and  some hypothesis on the data, the likelihood ratio converges almost surely to the information exchange rate $I_\infty( x_A \leftrightarrow x_B)$.

A related encouraging result due to \cite{PermKW11} is the emergence of the directed information in the false-alarm probability error rate.
Merging the two tests (\ref{test1:eq}),(\ref{test2:eq}), {\it i.e.,} testing both for causality and coupling, or neither, the test is written as:
\begin{eqnarray}
\left\{\begin{array}{lclcl}
H_0    & : &  p_0(x_B(i)\mid x_A^{i}, x_B^{i-1} )  &=& p( x_B(i)\mid x_B^{i-1} )  , \forall i\leq T \\
H_1     &:  & p_1(x_B(i)\mid x_A^{i}, x_B^{i-1} ) &=& p( x_B(i)\mid x_A^{i}, x_B^{i-1} ),  \forall i\leq T 
 \end{array}\right.
\end{eqnarray}
Among the tests with a probability of miss $P_M$ that is lower than some positive value $\varepsilon>0$, the best probability of false alarm $P_{FA}$ follows $\exp \big( -T  I(x_A\rightarrow x_B)\big)$ when  $T$  is large.  For the case studied here, this is the so-called Stein lemma \cite{CoveT06}.

In the multivariate case, there is no such  result in the literature. An  extension is proposed here. However, this is  restricted to the case of instantaneously {\em uncoupled} time series. Thus, we assume for the 
end of this subsection that:
\begin{eqnarray}
p( x_A(i),x_B(i),x_C(i)\mid x_A^{i-1}, x_B^{i-1},x_C^{i-1} ) =  \prod_{\alpha=A,B,C} p( x_\alpha(i)\mid x_A^{i-1},x_B^{i-1},x_C^{i-1} ) , \mbox{ }   \forall i\leq T
\end{eqnarray}
which means that there is no instantaneous exchange of information between the three subsets that form a partition of $V$.
This assumption has held in most of the recent studies that have applied Granger causality tests. It is, however, unrealistic in applications where the dynamics of the processes involved are faster than the sampling period adopted (see \cite{Gran80} for a discussion in econometry).
Consider now the problem of testing Granger causality of $A$ on $B$ relative to $V$. The binary hypothesis test
is given by:
\begin{eqnarray}
\left\{\begin{array}{lclcl}
H_0    & : & p_0(x_B(i)\mid x_A^{i-1}, x_B^{i-1},x_C^{i-1} )  &=& p( x_B(i)\mid x_B^{i-1},x_C^{i-1}  )  , \forall i\leq T \\
H_1     &:  & p_1(x_B(i)\mid x_A^{i-1}, x_B^{i-1},x_C^{i-1}  ) &=& p( x_B(i)\mid x_A^{i-1}, x_B^{i-1} ,x_C^{i-1} ),  \forall i\leq T 
 \end{array}\right.
\end{eqnarray}
The log likelihood {\it ratio } reads as:
\begin{eqnarray}
l(x_A^T,x_B^T,x_C^T) &=& \sum_{i=1}^T \log \frac{p( x_B(i)\mid x_A^{i-1}, x_B^{i-1} ,x_C^{i-1} )}{p( x_B(i)\mid x_B^{i-1},x_C^{i-1}  ) }
\end{eqnarray}
Again, by assuming that  the law of large numbers applies, we can conclude
that under $H_1$
\begin{eqnarray}
\frac{1}{T}l(x_A^T,x_B^T,x_C^T)   \xrightarrow{T\rightarrow +\infty} T_\infty( x_A \rightarrow x_B \| x_C)   \mbox{ a.s.}
\end{eqnarray}
This means that the causal conditional transfer entropy rate is the limit of the log likelihood {\it ratio }  as the amount of data increases.

\subsubsection{Prediction based approach in the Gaussian case}

Following definition 1 and focusing on  the quadratic risk  $R(e)=E[e^2]$, Geweke introduced the following indices for the study of Gaussian stationary processes \cite{Gewe82,Gewe84}:
\begin{eqnarray}
F_{x_A\leftrightarrow x_B}&=&\lim_{n\rightarrow +\infty}  \frac{R(x_B(n) | x_B^{n-1},x_A^{n-1})}{R(x_B(n) | x_B^{n-1},x_A^{n})}\\
F_{x_A\leftrightarrow x_B \| x_C}&=&\lim_{n\rightarrow +\infty} \frac{R(x_B(n) | x_B^{n-1},x_A^{n-1}, x_C^n)}{R(x_B(n) | x_B^{n-1},x_A^{n-1},x_C^n)}\\
F_{x_A\rightarrow x_B}&=&\lim_{n\rightarrow +\infty}  \frac{R(x_B(n) | x_B^{n-1})}{R(x_B(n) | x_B^{n-1},x_A^{n-1})} \\
F_{x_A\rightarrow x_B \| x_C}&=&\lim_{n\rightarrow +\infty} \frac{R(x_B(n) | x_B^{n-1}, x_C^{n-1})}{R(x_B(n) | x_B^{n-1},x_A^{n-1},x_C^{n-1})}
\end{eqnarray}
Geweke demonstrated the efficiency of these indices for testing Granger causality and instantaneous coupling (bivariate and multivariate cases). Furthermore, in the bivariate case, Geweke showed that:
\begin{eqnarray}
F_{x_A\rightarrow x_B} + F_{x_B\rightarrow x_A} + F_{x_A\leftrightarrow x_B}= I_\infty(x_A;x_B) 
\label{Gewedecomp:eq}
\end{eqnarray}
where $I_\infty(x_A;x_B)$ is the mutual information rate. This relationship that was already sketched out in \cite{Gran63}, is nothing but Eq. 
(\ref{sums2diinst:eq}). Indeed,  in the Gaussian case,  $F_{x_A\leftrightarrow x_B} = I_\infty(x_A\leftrightarrow x_B) $ and $F_{x_A\rightarrow x_B} = I_\infty(x_A\rightarrow x_B) $ stem from the knowledge that the entropy rate of a Gaussian stationary process is the logarithm of the asymptotic power of the one-step-ahead prediction \cite{CoveT06}.
Likewise, we can show that $F_{x_A\leftrightarrow x_B \| x_C} = I_\infty(x_A\leftrightarrow x_B \| x_C) $ and $F_{x_A\rightarrow x_B \| x_C} = I_\infty(x_A\rightarrow x_B \| x_C) $ holds. 

In the multivariate case, conditioning on  the past of the side information, {\it i.e.} $x_C^{n-1}$, in  the definition 
of $F_{x_A\leftrightarrow x_B \| x_C}$, a decomposition 
analagous to Eq. (\ref{Gewedecomp:eq}) holds, and is exactly that given by Eq. (\ref{sums2diinstcond:eq}).

\subsubsection{The model-based approach }

In a more general framework, we examine how  a model-based approach can be used to test for Granger causality, and how directed information comes into play.

Let us consider a rather general model in which $x_V(t)$ is a multivariate Markovian process that statisfies:
\begin{eqnarray}
x_V(t) = f_\theta\big( x_{V  t-k}^{\mbox{ } \,\,\,\, t-1} \big)  + w_V(t)
\end{eqnarray}
where $f_\theta: \R^{k|V|} \longrightarrow \R^{|V|}$ is a function belonging to some functional  class ${\cal F}$, and where $w_V$ is a multivariate i.i.d. sequence, the components of which are not necessarily mutually independent.
Function $ f_\theta$ might  (or might not) dependon $\theta$, a multidimensional parameter. This general model  includes as a particular case, linear multivariate autoregressive with moving average (ARMA)  models, and nonlinear ARMA models; $ f_\theta$ can also stand for a function belonging to some reproducing kernel Hilbert space, which can be estimated from the data \cite{SchoS02,MariPS08,AmblVMR12}.  
Using the partition $A,B,C$, this model can be written equivalently as:
\begin{eqnarray}
\left\{\begin{array}{lcl}
x_A (t)& =& f_{A,\theta_A}\big( x_{A  t-k}^{\mbox{ } \,\,\,\, t-1} ,x_{B  t-k}^{\mbox{ } \,\,\,\, t-1} ,x_{C  t-k}^{\mbox{ } \,\,\,\, t-1} \big)  + w_A(t) \\
x_B( t)& =& f_{B,\theta_B}\big( x_{A  t-k}^{\mbox{ } \,\,\,\, t
-1} ,x_{B  t-k}^{\mbox{ } \,\,\,\, t-1} ,x_{C  t-k}^{\mbox{ } \,\,\,\, t-1} \big)  + w_B(t) \\
x_C (t)& =& f_{C,\theta_C}\big( x_{A  t-k}^{\mbox{ } \,\,\,\, t-1} ,x_{B  t-k}^{\mbox{ } \,\,\,\, t-1} ,x_{C  t-k}^{\mbox{ } \,\,\,\, t-1} \big)  + w_C(t) 
\end{array}\right.
\end{eqnarray}
where the functions $f_{.,\theta_.} $ are the corresponding components  of $f_\theta$. 
This relation can be used  for inference in a parametric setting: the functional form is assumed to be known and the determination of the function is replaced by the estimation of the parameters $\theta_{A,B,C}$.  This can also be used in a nonparametric setting, in which case the function $f$ is searched for in an appropriate functional space, such as an rkHs associated to a kernel \cite{SchoS02}.

In any case, for studying the influence of $x_A$ to $x_B$ relative to $V$,  two models are required for $x_B$: one in which $x_B$ explicitly depends on $x_A$, and the other one in which  $x_B$ does not depend on $x_A$. In the parametric setting, the two models can be merged into a single model, in such a way that some components of the parameter $\theta_B$ are,  or not, zero, which dependis whether $A$ causes $B$ or not. The procedure then consists of testing nullity (or not) of these components.  
 In the linear Gaussian case, this leads to the Geweke indices discussed above.  In the nonlinear (nonGaussian) case, the Geweke indices can be used to evaluate the prediction in some classes of nonlinear models (in the minimum mean square error sense). In this latter case, the decomposition of the mutual information, 
Eq. (\ref{Gewedecomp:eq}), has no reason to remain valid. 

Another approach base relies on directly modeling the probability measures. This approach has been used recently to model spiking neurons and to infer Granger causality between several neurons working in the class of generalized linear models \cite{QuinCKH11,KimPGB11}.
Interestingly, the approach has been used either to estimate the directed information \cite{QuinCKH11} or to design a likelihood ratio test \cite{GourMR87,KimPGB11}. Suppose we wish to test whether '$x_A$ Granger causes $x_B$ relative to $V$'
as a binary hypothesis problem, as in section \ref{binaryhyp:ssec}. Forgetting the problem of instantaneous coupling, the problem is then to choose between the hypotheses:
\begin{eqnarray}
\left\{\begin{array}{lclcl}
H_0    & : & p_0(x_B(i)\mid x_V^{i-1} )  &=& p( x_B(i)\mid x_V^{i-1};  \theta_0 ) , \forall i\leq T \\
H_1     &:  & p_1(x_B(i)\mid x_V^{i-1} ) &=& p( x_B(i)\mid x_V^{i-1};  \theta_1 ),  \forall i\leq T 
 \end{array}\right.
 \label{test4:eq}
\end{eqnarray}
where the existence of causality  is entirely reflected into the parameter $\theta$.
To be more precise, $\theta_0$ should be seen as a restriction of $\theta_1$ when its components linked to $x_A$ are set to zero.
 As a simple example using the model approach discussed above,  consider the simple linear Gaussian model
 \begin{eqnarray}
x_B(t) = \sum_{i>0} \theta_A(i) x_A(t-i) +\sum_{i>0} \theta_B(i) x_B(t-i) +\sum_{i>0} \theta_C(i) x_C(t-i)  + w_B(t)
\end{eqnarray}
where $w_B(t)$ is an i.i.d. Gaussian sequence, and $\theta_A,\theta_B,\theta_C$ are multivariate impulse responses of appropriate dimensions. Define $\theta_1=(\theta_A,\theta_B,\theta_C)$ and $\theta_0=(0,\theta_B,\theta_C)$. Testing for Granger causality is then equivalent to testing $\theta=\theta_1$; furthermore, the likelihood {\it  ratio }  can be implemented due to the Gaussian assumption.  The example developed in \cite{QuinCKH11,KimPGB11}, 
assumes that the probability that  neuron $b$ ($b\cup A \cup C= V$)
sends a message at time $t$ ($x_b(t)=1$)  to its connected neighbors is given by the conditional probability
\begin{eqnarray*}
\mbox{Pr}\big( x_b(t)=1 \big| x_V^t; \theta \big)= U\big(  \sum_{i>0} \theta_A(i) x_A(t-i) +\sum_{i>0} \theta_b(i) x_b(t-i) +\sum_{i>0} \theta_{Eb}(i) x_{Eb}(t-i)  + w_b(t) \big) 
\end{eqnarray*}
where $U$ is some decision function, the output of which belongs to $[0 ; 1]$, $A$ represents the subset of neurons that can send information to $b$, and $Eb$ represents external inputs to $b$.  Defining this probability for all $b\in V$ completely specifies the behavior of the neural network $V$.

The problem is a  composite hypothesis testing problem, in which parameters defining the likelihoods have to be estimated. It is known that tere is no definitive answer to this problem \cite{LehmR05}. An approach that relies on an estimation of the parameters using maximum likelihood can be  used. Letting $\Omega$ be the space where parameter $\theta$ is searched for and $\Omega_0$ the subspace where $\theta_0$ lives, then
  the generalized loglikelihood {\it ratio } test reads:
\begin{eqnarray}
l(x_A^T,x_B^T) &:=& \log \frac{\sup_{\theta \in \Omega} p(x_V^T ; \theta)}{\sup_{\theta \in \Omega_0} p(x_V^T ; \theta)} = \log \frac{ p(x_V^T ; \widehat{\theta^T_1})}{ p(x_V^T ; \widehat{\theta^T_0})}
\end{eqnarray}
where  $\widehat{\theta^T_i}$  denotes the maximum likelihood estimator of $\theta$ under hypothesis $i$.
 In the linear Gaussian case, we will recover exactly the measures developed by Geweke. In a more general case, 
and as illustrated in section \ref{binaryhyp:ssec}, as the the maximum likelihood estimates are efficient, we can conjecture that the generalized log likelihood {\it ratio} will converge to the causal conditional  transfer entropy rate
if sufficiently relevant conditions are imposed on the models ({\it e.g.,} Markov processes with recurrent properties).
This approach  was described in \cite{GourMR87} in the bivariate case.

%%%%%%%%%%%%%%%%%%%%%%%%%%%%%%%%%%%%%%%%%%%%%%%%%%%%%%%%%%%%

\section{Conclusions}

 Granger causality was developed originally in econometrics, and it is now transdisciplinary, 
with the literature on the subject is widely dispersed. We have tried here to sum up the profound links that exist between Granger causality and directed information theory. The key ingredients to build these links are conditional independence and the recently introduced causal conditioning.

We have eluded the important question of how to practically use the definitions and measures presented here. Some of the measures can be used and implemented easily, especially in the linear Gaussian case. In a more general case, different approaches can be taken.  The information-theoretic measures can be estimated, or the prediction can be explicitly carried out and the residuals used to assess causality. 

Many studies have been carried out over the last 20 years on the problem of estimation of information-theoretic measures. We refer to  \cite{KozaL87,BeirDGM97,Pani03,KrasSG04,GoriLMN05} for information on the different ways to estimate information measures. Recent studies into the estimation of entropy and/or information measures are \cite{LeonPS08, WangKV09,SricRH12}. The recent report by \cite{ViceWLP11} extensively details and applies transfer entropy in neuroscience  using $k$-nearest neighbors type of estimators.
Concerning the applications,  important reviews include \cite{HlavSPVB07,Palu07}, where some of the ideas discussed here are also mentioned, and where practicalities such as the use of surrogate data, for example, are extensively discussed. Applications  for neuroscience are discussed in \cite{KamiDTB01,GourBF06,KimPGB11,Eich05,Eich06}.

Information-theoretic measures of conditional independence based on Kullback divergence were chosen here to 
illustrate the links between Granger causality and (usual) directed information theory. Other type of divergence  could have been chosen; metrics in probability space could also be useful in the assessing of conditional independence. 
As an illustration, we refer to the study of Fukumizu and co-workers \cite{FukuGSS07}, where conditional independence was evaluated using  the Hilbert-Schmidt norm of an operator between reproducing kernel Hilbert spaces. The operator generalizes the partial covariance between two random vectors given a third one, and is called the conditional covariance operator. Furthermore,  the Hilbert-Schmidt norm of conditional covariance operator can be efficiently estimated from data. A related approach is also detailed in   \cite{SethP12}.

Many important directions can be followed. Causality between nonstationary processes has  rarely been considered (see however  \cite{ViceWLP11} for an {\it ad-hoc } approach in neuroscience). A very  promising methodology is to adopt a graphical modeling way of thinking. The result of \cite{Eich11} on the structural properties of Markov-Granger causality graphs can be used to identify such graphs from real datasets.  A first step in this direction was proposed by 
\cite{QuinKC11}. Assuming that the network under study is a network of sparsely connected nodes and that some Markov properties hold, efficient estimation procedures can be designed, as is the case in usual graphical modeling. 

%%%%%%%%%%%%%%%%%%%%%%%%%%%%%%%%%%%%%%%%%%%%%%%%%%%%%%%%%%%%

\section*{Acknowledgements}
P.O.A. is supported by a Marie Curie International Outgoing Fellowship from the European Community.

%==========================================================
%==========================================================
% Back Matter (References and Notes)
%----------------------------------------------------------
% Style and layout of the references
\bibliographystyle{mdpi}
\makeatletter
\renewcommand\@biblabel[1]{#1. }
\makeatother
%----------------------------------------------------------
% Use the following option to include external BibTeX files:
%\bibliography{template}
%----------------------------------------------------------

\end{document}